\documentclass[
  reprint,
  prd,
  amsmath,
  amssymb,
]{revtex4-2}
\usepackage[normalem]{ulem}
\usepackage{graphicx}% Include figure files
\usepackage{dcolumn}% Align table columns on decimal point
\usepackage{bm}% bold math
\usepackage{hyperref}
\usepackage[table]{xcolor}
\hypersetup{
    colorlinks,
    linkcolor={red},
    citecolor={blue},
    urlcolor={blue}
}
\usepackage{booktabs} 
\usepackage{xcolor}
\newcommand{\blue}[1]{\textcolor{blue}{#1}}

\usepackage{amsthm}

\newtheorem{theorem}{Theorem}

\usepackage[utf8]{inputenc}
\begin{document}

%\preprint{APS/123-QED}

%\title{Constraining inflationary models based on de-Sitterization Bound}% Force line breaks with \\

\title{Constraining models of inflation via an asymptotic de Sitter realization of Bianchi Cosmologies}

% \title{Constraining inflationary models via de-Sitterization of Bianchi Cosmologies} 
% cosmic no hair conjecture}

\author{Apurba Samanta}
\email{apurbasamanta79@gmail.com}
\author{Rahul Kothari}%
\email{quantummechanicskothari@gmail.com}
\affiliation{School of Physical Sciences, Indian Institute of Technology Mandi,  Himachal Pradesh, 175005, India
}%


% \affiliation{
%  Second institution and/or address\\
%  This line break forced% with \\
% }%
% \affiliation{
%  Third institution, the second for Charlie Author
% \affiliation{%
%  Authors' institution and/or address\\
%  This line break forced with \textbackslash\textbackslash
% }%

% \collaboration{CLEO Collaboration}%\noaffiliation


\begin{abstract}
Universe is isotropic and homogeneous at $\gtrsim 100$ Mpc length scales. 
It is desirable that  present state of the Universe has no dependence on its initial geometry. In case of Bianchi Universes, i.e., anisotropic but homogeneous Universe, this has already been demonstrated via cosmological constant in a process that we call \textit{de Sitterization}. In this paper, we show that for Bianchi Universe, the same state can be achieved by a homogeneous inflaton field with a general potential and satisfying a criterion, without needing cosmological constant. More importantly, we show that the same condition can constrain models of inflation and explain our idea with  examples. Our results indicate that those model of inflation are favoured by observations which wipe out initial anisotropy. 
\end{abstract}

\maketitle

%\tableofcontents

\section{\label{sec:intro}Introduction}
Inflation \cite{PhysRevD.23.347, LINDE1982389, Linde:1983gd, Linde_1994, Starobinsky:1980te,Lemoine:2008zz}  is an important epoch in the history of the Universe during which it went through an exponential expansion. This period of rapid expansion, on top of standard big bang cosmology, is needed to resolve many puzzles like horizon problem, flatness problem, etc. Additionally, inflation is considered to be the source of  small fluctuations leading to CMB anisotropies and eventually large scale structure. It turns out that many models of inflation are consistent with the observations \cite{2020,2014,2016,Martin_2014,Martin:2024qnn,MARTIN2024101653,IJJAS2013261}. %Recent efforts include numerical parameter estimation via ModeCode \cite{Mortonson_2011,caravano2025inflationeasyclatticecode}, restrictions on warm inflation scenarios \cite{PhysRevD.107.063543}, and related model evaluations \cite{PhysRevD.104.103528,li2025constraintscanonicalsinglefieldslowroll,LEON2019100285,Martin_2014,Martin:2024qnn}.

While the issues of primordial geometry and matter distribution of the Universe remain fundamentally unresolved, the cosmic no-hair conjecture (CNHC) \cite{PhysRevD.15.2738, HAWKING198235} posits that the late-time geometry of an expanding universe becomes largely insensitive to its initial anisotropic geometry. Recent constraints on cosmological anisotropy strongly support CNHC \cite{PhysRevLett.117.131302}. Related questions have also been investigated in the context of Kerr–Schild geometries of black holes \cite{Sharma_2022}  In other words, it asks the question whether our Universe, irrespective of the initial geometry (homogeneous or not, isotropic or not, etc.),  eventually evolves to the de Sitter state after a few e-folds of inflation.  In this context, Refs. \cite{PhysRevD.28.2118,  BarrowGotz1983,Barrow:1990td,Nakao:1990vq,Heusler:1991ep} show that homogeneous spaces, referred to as  \textit{Bianchi Universes} (apart from type IX), with a cosmological constant, evolve towards de Sitter state. Similar results are also available for inhomogeneous cosmologies \cite{PhysRevD.35.1146,Jensen:1986vs,PhysRevD.35.2345,Starobinsky1983}, at least locally. Nonetheless, CNHC breaks down in various scenarios, when energy conditions are relaxed or  matter is modelled strictly as a perfect fluid \cite{barrow1987cosmic,PhysRevD.61.064012,PhysRevD.85.123508,Sarkar2025,PhysRevLett.102.151302,Kanno:2010nr}, in the context of dilaton coupling in $R^2$ gravity \cite{Maeda:1987xf},   persistent spatial anisotropies  \cite{Yamamoto_2012,Soda_2012}, inclusion of exotic multi-field dynamics and Vector filed \cite{Do:2011zza,Do_2021,PhysRevLett.102.191302}, etc. 
%\red{[ A robust body of literature demonstrates CNHC failure across diverse theoretical frameworks. It is well established that the CNHC breaks down when Wald’s . Furthermore, exact solutions featuring positive spatial curvature and a positive cosmological constant reveal ever-expanding closed universes that completely avoid a de Sitter phase, challenging the standard assumption that $\Lambda$-driven expansion necessitates inflation \cite{Gotz:1988xm,PoncedeLeon:1988uqh}. Modifications to gravity also restrict the conjecture's domain; for instance, the introduction of a dilaton coupling in $R^2$ gravity eliminates inflationary behavior entirely \cite{Maeda:1987xf}. Persistent spatial anisotropies provide another definitive class of counterexamples, particularly in models where scalar inflation drivers non-minimally couple to vector fields to stably generate vector hair \cite{Yamamoto_2012,Soda_2012}. This anisotropic inflation permanently violates the no-hair theorem and leaves observable statistical anisotropies in primordial fluctuations \cite{Soda_2012}. Finally, the inclusion of exotic multi-field dynamics, such as phantom fields in Bianchi type I cosmologies, actively prevents isotropization by inducing instabilities, demonstrating that early-universe fields frequently sustain anisotropy rather than uniformly diluting it \cite{Do:2011zza,Do_2021}.]}

It turns out that an exact realization of CNHC, carried out by a cosmological constant isn't observationally viable, as it precludes a graceful exit {out of a perpetual de Sitter evolution}. A viable inflationary model must continue for {at least} 55–60 e-folds of expansion to be consistent with CMB observations. {We can tackle both issues of initial geometry and  graceful exit out of inflation by} %To achieve this, we can 
dividing inflationary period into two distinct phases:  initial exact de Sitter and  subsequent quasi-de Sitter phases. While the quasi-de Sitter epoch is well-studied and successfully modelled by various scalar field potentials such as exponential, power-law, or polynomial forms \cite{Martin_2014} the initial exact de Sitter phase is rarely described in detail. Many studies assume a starting exponential expansion driven by an effectively constant potential. In this regard, relaxing constraints on initial geometries highlights the crucial role of \textit{de-Sitterization}. It serves as the primary mechanism towards achieving isotropy and homogeneity in the early Universe. %, successfully bridging arbitrary initial conditions with current observational constraints. 
%{Although observationally safeguarding the electroweak vacuum from inflationary de Sitter fluctuations \cite{Espinosa_2008} requires a low-energy scale, confining the paradigm to the exact plateau potentials plagued by initial conditions, multiverse, and ``unlikeliness'' problems.

In this paper, we present a framework that relaxes this geometric prior to an initially anisotropic Bianchi background. %Consequently, to reconcile CNHC with observations, inflation must be driven by dynamically evolving scalar field potential rather than an idealized constant state \cite{kitada1992cosmic,chakraborty2001inflation, Burd:1988ss,Moss:1986ud,Bradas:1987mk,Maeda:1988rb,Barrow:1990td}. In other words, a realistic model of inflation must rely on a quasi-de Sitter framework rather than an exact de Sitter spacetime \cite{PhysRevD.107.023527, Bhatt2024InflationDynamics, mu2026inflationdrivenbarecosmological}. %Conceptually, the inflationary period can be divided into two distinct epochs: . 
Working with a general potential function $V(\phi) = V_0 + f(\phi)$ of single scalar field, where $V_0$ is constant and $f(\phi)$ is $\phi$ dependent part, we demonstrate that 
\begin{enumerate}
\item For any given potential $V(\phi)$, an initial Bianchi geometry (except type IX), asymptotically achieves de Sitter state through $V_0$ without the need of cosmological constant, thereby generalizing previous results  \cite{kitada1992cosmic, chakraborty2001inflation}. %A well-defined parameter space for the potential coefficients enforces an initial exact \blue{[approximate?]} de Sitter phase. 
Additionally, $f(\phi)$  drives quasi de-Sitterization, aiding in achieving a graceful exit out of otherwise perpetual de Sitter evolution %This dynamically evolves into a quasi-de Sitter regime and guarantees a graceful. % exit within a single, self-consistent picture.
 %, driven by a  homogeneous scalar field with a potential  
given that \textit{de Sitterization condition}  gets satisfied. %asymptotically achieves the de Sitter state if it satisfies \textit{de Sitterization condition} , without requiring  cosmological constant. We show this to be true without assuming any specific  $V(\phi)$, .   %We show that the constant part of the potential $V_0$ acts like a cosmological constant. 
\item %We rephrase the {de Sitterization condition} in  terms of \textit{de Sitterization parameter} $\alpha$ and show that this
The de Sitterization parameter $\alpha$ can be used to constrain models of inflation. We find that models with $\alpha>1$ are ruled out by observations. %Standard single-field plateau models typically invoke isotropic FLRW initial conditions to satisfy Planck constraints on $n_\mathrm{s}$ and $r$ via quasi-de Sitter expansion. 
\end{enumerate}
In this study, we restrict ourselves to Bianchi Universes and leave inhomogeneous cosmologies for a future study. We work in Planck mass $M_\mathrm{Pl}$ units with $M_\mathrm{Pl}=1$ that renders both potential $V(\phi)$ and scalar field $\phi$ dimensionless. %Utilizing the Theorem~[\ref{thm:extremal-Chebyshev-Property}, we derive a unique parameter bound for this de Sitterization mechanism. By encapsulating this bound and the necessary initial conditions into a single parameter $\alpha$, we systematically evaluate known inflationary models. Addressing the unresolved initial condition for inflation \cite{Linde:2017pwt,Dimopoulos:2016yep,Mishra:2018dtg,Artymowski:2016pjz,Ijjas_2013,Brandenberger:2016uzh}, our bound indirectly ensures an initial de Sitter phase of some $e$-folds. Crucially, it shows that potentials lacking an initial constant-like state trigger a premature graceful exit before the required 55-60 $e$-folds, precluding successful inflation. Furthermore, confronting our framework with Planck data reveals that observationally excluded models consistently yield larger values of $\alpha$. 
 %This confirms that potentials lacking an initial constant-like regime are fundamentally nonviable. Crucially, our approach establishes $\alpha$ as a novel, single-parameter diagnostic for evaluating inflationary models.

%We set $c=1$ and express potential and scalar field in units of Planck length $M_\mathrm{Pl}=1$. This renders scalar field $\phi$ and potential $V(\phi)$ as dimensionless quantities.

%\blue{We also define inflation that comprises of (a) de Sitter and (b) quasi de Sitter era.}
% Begin with somewhat very broad survey of the term inflation -- why it was introduced, what does it do. That there are a myriad of models that explain the observations. Here in the article, we propose a method to constrain the models in a specific manner.

\section{De-Sitterization with scalar field potential\label{sec:de-Sitter-with-potential}}
We consider inflation potential to be an analytic function of $\phi$ 
\begin{equation}\label{eq:modified potential form}
V(\phi)=V_{0}+f(\phi)
\end{equation}
where $V_0$ is constant and $f(\phi)$ contains $\phi$, $\phi^2$, etc., terms. With  $V_0$ alone, Universe can achieve a de Sitter state, since it acts like a cosmological constant. But by itself this is insufficient because Universe undergoes perpetual inflation and won't have any graceful exit. Thus we need  $f(\phi)\ne 0$, albeit small as compared to $V_0$. We can state this condition in terms of \textit{de-Sitterization parameter} $\alpha$. 
\begin{equation}
    |F(\phi)| \le \alpha,\ F(\phi):= \begin{cases}
        f(\phi)/V_0 & V_0 \ne 0 \\
        f(\phi) & V_0 =0
    \end{cases}\label{eq:de-sitter-condition}
\end{equation}
A constant potential corresponds to pure de Sitter evolution which can happen when $\alpha=0$ for $V_0\ne 0$ case. Additionally, for $V_0=0$, $\alpha=0$ implies unphysical scenario of having $V(\phi)=0$.
Thus we enforce $\alpha>0$.
%From this we can see that $\alpha=0$, forces $F(\phi)=0$ thereby implying only a perpetual de Sitter evolution, so we need  $\alpha>0$. 
We also impose   $\alpha< 1$ which ensures the realization of the de Sitter state, as discussed in \S\ref{sec:bounds-on-parametrs}.
% a later section}. 
% \red{[No,not agree with this]}%ensures that the de Sitterization process occurs within a specific inflationary regime. Beyond this range, additional fine-tuning of the potential is required to achieve a graceful exit after 55--60 e-folds. 
{Summarily, we state the de Sitterization condition as 
\begin{equation}
    0<\alpha<1 \label{eq:alpha-cond}
\end{equation}
While standard scenarios assume initial isotropy, we consider an initially anisotropic but homogeneous spacetime also known as Bianchi spacetime. Thus, the bound \eqref{eq:alpha-cond} is required to sustain the de-Sitterisation era, driving the decay of early-time anisotropy. {This condition ensures that approximately isotropization is completed before the onset of the system's graceful exit.}
% This constraint secures a stable quasi-de Sitter intermediate phase necessary for complete isotropization 
% \blue{[but isotropization has already been achieved!]}\red{[agree with this]} before the system undergoes a graceful exit. 
% then a successful quasi de Sitter epoch ensues after a de Sitter epoch. %Since inflationary regime consists of both these epochs, 
% It follows that a successful de Sitter epoch together with \eqref{eq:alpha-cond} helps achieving inflation within $55-60$ e-folds.
As a bonus, the same condition constrains inflationary models with just one parameter. We give an example of estimating $\alpha$ in \S\ref{sec:example}.
}

%Before demonstrating how we can achieve the de Sitter state from a potential $V(\phi)$ satisfying \eqref{eq:de-sitter-condition}, we explicitly state an assumption used in this letter
% \begin{equation}
%     \text{de Sitter}\Rightarrow \text{quasi de Sitter}
% \end{equation}
% i.e., if Universe undergoes de Sitterization, then it also achieves quasi de Sitter state. Further, since inflation consists of both de Sitter and quasi de Sitter epochs, it follows that
% \begin{equation}
%     \text{de Sitter}\Rightarrow \text{inflation}
% \end{equation}
% Thus, if the Universe achieves de Sitter state, it also achieves successful inflation. 
%Now we describe how we can achieve de Sitter state with any given potential $V(\phi)$ satisfying \eqref{eq:alpha-cond}.
%In this section, we demonstrate that both class A and B Bianchi cosmologies (except Bianchi IX) asymptotically approach a de Sitter state, process that we call ``de Sitterization'' in this paper. %We should emphasize that our Since we demonstrate that with the potential $V(\phi)$ only Bianchi cosmologies de Sitterize, we can't just extrapolate that the same will be true for inhomogeneous case, unless we prove it. Unlike previous treatments relying on an explicit cosmological constant \cite{PhysRevD.28.2118}, 
\subsection{Achieving the de Sitter state using $V(\phi)$}
Ref \cite{PhysRevD.28.2118} considers a cosmological constant $\Lambda$ without making any further assumptions regarding the energy momentum tensor used in the Einstein equations. %In our %minimalistic framework, constant part of the potential acts like a cosmological constant and drives de Sitter state while  $f(\phi)$ achieves the quasi de Sitter epoch. 
We consider scalar field and matter energy momentum tensor in the Einstein field equations 
\begin{equation}\label{eq:Einstein-Equation}
    R_{ab} - \frac{1}{2} R g_{ab}  = \kappa^{2} \left[T_{ab}^{(M)}+T_{ab}^{(\phi)}\right],\ \kappa^{2}=8\pi G
\end{equation}
%introduces a more realistic extension by replacing $\Lambda$ with a scalar field $\phi(t)$ featuring a general potential Eq.~\eqref{eq:modified potential form}. 
We don't make any assumptions regarding the matter energy-momentum tensor $T^{(M)}_{ab}$ but  assume that it satisfies weak and strong energy conditions.
\begin{align}
    \text{WEC} &: T^{(M)}_{ab}n^{a}n^{b} \geq 0, \label{eq:WEC} \\
    \text{SEC} &:
    \left(T^{(M)}_{ab}-\frac{1}{2}g_{ab}T^{(M)}\right)n^{a}n^{b} \geq 0
\label{eq:SEC}
\end{align}
where $n^{a}$ is a timelike unit vector. On the other hand, the scalar field energy momentum tensor $T^{(\phi)}_{ab}$ is assumed to be homogeneous and isotropic that obeys WEC but violates the SEC. %depending solely on cosmic time. Consequently, the matter sector $T^{(M)}_{\mu\nu}$ naturally satisfies the , whereas the scalar field $T^{(\phi)}_{\mu\nu}$ obeys the WEC but violates the SEC under the following conditions: The corresponding Einstein field equations are To demonstrate Wald's theorem, we adopt all Class A and B Bianchi cosmologies, with the explicit exception of Bianchi IX as our background. Utilizing ordinary matter and a homogeneous scalar field with an arbitrary potential V($\phi$) as the source, the Einstein field equations are derived from the action \cite{kitada1992cosmic}:
Next, we utilize Gauss-Codazzi relations \cite{gourgoulhon200731formalismbasesnumerical, wald1984general} to obtain 
\begin{align}
    \frac{K^{2}}{3}&=\kappa^{2}\left(\frac{1}{2}\dot{\phi}^{2}+V(\phi)\right)
+\frac{1}{2}\left(\sigma_{ab}\sigma^{ab}
-R^{(3)}\right)\notag\\
&\quad+\kappa^{2}T^{(M)}_{ab}u^{a}u^{b}. \label{eq:Guass-Codazzi-simpli}
\end{align}
We assume initially expanding universe, so that $K>0$ \cite{kitada1992cosmic,chakraborty2001inflation}. Further using, Raychaudhuri equation \cite{PhysRev.98.1123} and standard kinematic properties, we get 
\begin{align}
\dot{K}&=-\frac{1}{3}K^{2}-\sigma_{ab}\sigma^{ab}+ \kappa^{2}(-\dot{\phi}^{2}+ V(\phi))\notag\\
        &\quad -\kappa^{2}\left(T^{(M)}_{ab}-\frac{1}{2}T^{(M)}g_{ab}\right)u^{a}u^{b} \label{eq:modified-raychau-cond}
\end{align}
% \blue{difference arises that we are using $V(\phi)$ in place of $\Lambda$.}
In these equations, \( u^{a} \) %as the unit normal vector emphasizes its special geometric and physical role. It 
represents the 4-velocity of comoving observers, who remain at fixed spatial coordinates and move only with the cosmic expansion.  Overdots denote derivative with respect to cosmic time $t$, $R^{(3)}$ is the 3D intrinsic spatial curvature, while the extrinsic curvature $K_{ab}$ features a trace-free anisotropic component of shear tensor $\sigma_{ab}$ that being purely spatial, satisfies \cite{poisson2004relativist}
\begin{equation}
    \sigma_{ab}\sigma^{ab}\ge0 \label{eq:sigma-condition}
\end{equation}
  and a trace $K = h^{ab}K_{ab}$, defined via the inverse spatial metric $h^{ab}$. 

Till this point, the analysis was general. To proceed further, we assume initial geometry as Bianchi. For Bianchi cosmologies, let the Killing vectors $X_a$ satisfy the commutation relation $[X_a,X_b] = C^{d}{}_{ab}X_d$. Then %starting with the Killing vectors $X_a$ that generate the underlying symmetry via the commutation relations $[X_a,X_b] = C^{d}{}_{ab}X_d$, Following Wald \cite{wald1984general}, we can express the system in terms of the structure constants $C^{d}{}_{ab}$ as
we can express the structure constants as \cite{wald1984general}
\begin{equation}\label{eq:structure constants}
    C^{c}_{\ ab} =M^{cd}\epsilon_{dab}+\delta^{c}_{[a}A_{b]}
\end{equation}
Here $M^{ab}$ is a symmetric tensor, and $\epsilon^{fdc}$ represents a three-form on the Lie algebra. The 3D Ricci scalar then reads
\begin{equation}\label{eq:3D-ricci-scalar}
    R^{(3)} = -\frac{3}{2}A_{b}A^{b} - h^{-1}\left(M^{ab}M_{ab} - \frac{1}{2}M^{2}\right),
\end{equation}
where, $h = \det(h_{ab})$, and $M=M^a_{\ a}$, the trace of $M$. Incorporating the Jacobi constraint $M^{ab}A_b = 0$ restricts the spatial curvature $R^{(3)} \leq 0$ for both Class A ($A_b = 0$) and Class B ($A_b \neq 0$), except Bianchi type IX. Notice that \eqref{eq:Guass-Codazzi-simpli} only makes sense if RHS is positive. Together with (i) $T_{\mu\nu}^{(M)}$ satisfying WEC, i.e., \eqref{eq:WEC} and (ii) condition \eqref{eq:sigma-condition}, this is guaranteed if $R^{(3)} \leq 0$, a condition that has been considered in previous papers. We believe that the case $R^{(3)} > 0$ also works if other terms are enough positive. However, at this point, we aren't able to give an explicit demonstration and ramifications of this conclusion. 

%Applying  WEC \eqref{eq:SEC} to \eqref{eq:Guass-Codazzi-simpli} for Bianchi models with $R^{(3)} \leq 0$, we deduce that $K^{2} \geq 0$. Assuming an initially positive potential $V(\phi) > 0$, this guarantees an initially expanding  ($K > 0$), analogous to Wald's demonstration. Furthermore, this framework can be generalized to Bianchi type-IX cosmologies, where $R^{(3)} \geq 0$. In this regime, the expansion condition $K^{2} \geq 0$ remains satisfied provided that the remaining terms contributions dominate over the positive spatial curvature $R^{(3)}$.  A generalized treatment of type IX would require demonstrating the dominance of the remaining terms over the positive three-curvature, which is not pursued in this analysis.

We can now derive the dynamical equation for the trace $K$ of extrinsic curvature. Using (i) SEC, i.e.,  \eqref{eq:SEC} in \eqref{eq:modified-raychau-cond} and condition  $\dot{\phi}^2 \ll V(\phi)$ valid through out the regime 
% (b) slow roll condition due to which $\dot{\phi}^2 \ll V(\phi)$ and 
% (b) we know that the $K^{2}>\kappa^{2}\left(\frac{1}{2}\dot{\phi}^{2}+V(\phi)\right)$ from \cite{chakraborty2001inflation} for all time  and
(ii) condition \eqref{eq:sigma-condition}, we get %Using \eqref{eq:Guass-Codazzi-simpli}, \eqref{eq:SEC} in \eqref{eq:modified-raychau-cond} and $R^{(3)} \leq 0$, we conclude that $K \geq 0$. This establishes a lower bound $K^{2} \geq 3\kappa^{2}V(\phi)$ on $K^2$, which subsequently enforces $\dot{K} \leq 0$. Also, {from \eqref{eq:modified-raychau-cond}, it follows that $\sigma_{ab}\sigma^{ab} \geq 0$ \cite{poisson2004relativist}. Employing the fact that for an expanding Universe $V(\phi) > 0$ and ordinary matter satisfies SEC (2nd inequality in \eqref{eq:SEC}), we get}
\begin{equation}\label{eq:fin-diff-equa}
\dot{K}\leq \kappa^{2}V(\phi)-\frac{1}{3}K^{2}
\end{equation}
We solve \eqref{eq:fin-diff-equa} perturbatively (see Appendix \ref{sec: linear}) after which 
% The resulting upper and lower bounds on \(K\) are given in \eqref{eq:total solution} and \eqref{eq:below range}, respectively,  
the upper and lower bounds on $K$ given in  \eqref{eq:total solution} and \eqref{eq:below range} can be compactly expressed as:
\begin{equation}\label{eq:compl_bound_K}
3\zeta \left(1 + \frac{\alpha}{2}\right) \le K \le \frac{3\zeta}{\tanh(\zeta t)}
+\frac{3\alpha}{2} \left[\frac{\zeta}{\tanh(\zeta t)}
-\frac{t\zeta^2}{\sinh^{2}(\zeta t)}\right]
\end{equation}
In the limit, \(t\gg 1/\zeta\), these bounds \eqref{eq:compl_bound_K} simplify,
% \red{\begin{equation}
%     3\zeta+\frac{\alpha\kappa^{2}V_{0}}{2\zeta} \le K \le 3\zeta+\frac{\alpha\kappa^{2}V_{0}}{2\zeta}
% \end{equation}}}
yielding a narrow, or ``squeezed,'' allowed interval for \(K\). Thus, in this limit
 % Combining these results, we obtain the following allowed range for K:
% \begin{widetext}
% \begin{align}\label{eq:upper and lower bound}
% 3\zeta+\frac{\alpha\kappa^{2}V_{0}}{2\zeta}\leq K\leq \frac{3\zeta}{\tanh(\zeta t)}+\frac{\alpha}{2} \left[\frac{\kappa^2 V_0}{\zeta\, \tanh(\zeta t)}
% -\frac{t\kappa^2 V_0}{\sinh^{2}(\zeta t)}\right]
% \end{align}
% \end{widetext}
% where $\zeta= \kappa\sqrt{{V_{0}}/{3}}$. 
% If we consider time large scales limit $t \gg 1/\zeta$, then \eqref{eq:repeat-K-solution} simplifies to %, since $\zeta t$ is dimensionless, $1/\zeta$ represents the characteristic time scale. We can therefore consider the long-time limit, .
% \red{We now examine the long-time limit of Eq.~\eqref{eq:repeat-K-solution}, $t\to\infty$, where $t$ is assumed to be much larger than the period of the $\phi_{0}\to\phi_{e}$ evolution illustrated in Fig.~\ref{fig:Potential_analogy}. }
\begin{equation}\label{eq:large_limit_solu}
 K\to 3\zeta \left(1 + \frac{\alpha}{2}\right).
\end{equation}
% Substituting the result of Eq.~\eqref{eq:large_limit_solu} into Eq.~\eqref{eq:Guass-Codazzi-simpli}, we find that, in the large-time limit, the matter contribution satisfies the dominant energy condition and the spatial curvature becomes negligible. Thus, the shear term is the leading contribution, giving
% \begin{equation}
% frac{ \sigma_{ab}\sigma^{ab}}{2}\to \frac{K^{2}}{3}-\kappa^{2}V(\phi).
% \end{equation}
% In this limit, $V(\phi)\simeq V_{0}(1+\alpha)$, while $K$ is given by Eq.~\eqref{eq:large_limit_solu}. Therefore,
% \begin{align}
% \frac{\sigma_{ab}\sigma^{ab}}{2}
% &\to \kappa^{2}V_{0}
% +\alpha\kappa^{2}V_{0}
% +\frac{\alpha^{2}\kappa^{2}V_{0}}{4}
% -\kappa^{2}V_{0}
% -\alpha\kappa^{2}V_{0} \notag\\
% &= \frac{\alpha^{2}\kappa^{2}V_{0}}{4}
% \to 0,
% \end{align}
% From \eqref{eq:alpha-cond}, the higher-order $\alpha^{2}$ contribution can be neglected, yielding $\sigma_{ab}\to0$. 
Using a similar argument, from \eqref{eq:upperbound of sigma} we conclude the anisotropy \(\sigma_{ab}\to0\) in the limit $t \gg 1/\zeta$. This indicates that spacetime approaches an isotropic de Sitter state at late times.  %Consequently, the anisotropic degrees of freedom are dynamically suppressed, leading to an asymptotically isotropic geometry. 
Starting from the general evolution equation presented in Ref. \cite{BarrowGotz1983}, the decay of the anisotropic contribution allows the evolution equation to be reduced to the simpler form \(\dot{h}_{ab}=2K_{ab}\). Integrating this equation and using {$t\gg 1/\zeta$} limit of \eqref{eq:large_limit_solu}, we obtain
%Applying this to \eqref{eq:total solution}, %and using the asymptotic relations
% \begin{equation}
% \sinh^{2}(\zeta t)\rightarrow\infty,
% \qquad
% \frac{\cosh(\zeta t)}{\sinh(\zeta t)}\rightarrow1
% \end{equation}
% Evaluating the time behaviour $t \to \infty$ yields a vanishing anisotropy $\sigma_{ab} \to 0$ and $K_{0}(t) \to 3\zeta$. Together with \eqref{eq:fin-diff-equa}, with the help of evolution equation $\dot{h}_{ab}=2K_{ab}$, we finally get
% \begin{equation} \label{eq:de~Sitterization}
%   h_{ab}(t)=h_{ab}(t_{0})\,e^{2\zeta(t-t_{0})}.
% \end{equation}
% . in the asymptotic limit, we obtain
\begin{equation}\label{eq:general geometry form}
h_{ab}(t)=h_{ab}(t_{0})\,\exp\left[2\zeta\left(1+\frac{\alpha}{2}\right)(t-t_{0})\right].
\end{equation}
This equation exhibits a perturbative deviation from the pure de Sitter behaviour. Further, when we take $\alpha\to0$ we recover results of \cite{Kitada:1992uh, PhysRevD.28.2118, chakraborty2001inflation}. While the condition~\eqref{eq:de-sitter-condition} reduces the generalized potential to an approximately constant form, the parameter $\alpha$ introduces a small correction to the asymptotic expansion rate. Consequently, for $\alpha <1$, the spacetime remains  asymptotically de Sitter up to sub-leading corrections. % of Eq.~\eqref{eq:de~Sitterization}. Equation~\eqref{eq:de~Sitterization} shows that, under the condition~\eqref{eq:de-sitter-condition}, 
Thus, our analysis shows that Bianchi cosmologies with a general potential, satisfying appropriate conditions, asymptotically de-Sitterize.
\section{Constraining Inflationary Models}
In this section, we first discuss slow roll parameters based method   for constraining the models of inflation. We then discuss our method based on de Sitterization condition.

\subsection{Slow roll parameters based constraints}
Under this paradigm, the inflation is described using various slow roll parameters given in terms of various order derivatives of the inflation potential $V(\phi)$ 
\begin{equation}
\begin{aligned}
\epsilon_V &= \frac{1}{2}\left(\frac{V'}{V}\right)^2 < 0.0048,
&
\eta_V &= \frac{V''}{V} = -0.0082 \\
\xi_V^2 &= \frac{V'V'''}{V^2} = -0.004,
&
\bar{\omega}_V^3 &= \frac{V'^2V''''}{V^3} = 0.0048
\end{aligned}
\label{eq:slow-roll-params-value}
\end{equation}
where the numerical values are taken from \cite{2020}.
%Slow rolling happens when the parameter values are small. 
Once the parameter $\epsilon_V$ reaches $1$, inflation stops and the epoch of reheating begins. Inflationary models have been constrained by mapping slow-roll parameters %(obtained using potential) 
with late-time observables --- scalar spectral index
$n_{s}$, the tensor-to-scalar ratio $r$ and running of the scalar spectral index $\alpha_{s}$  \cite{PhysRevD.65.101301, 10.1093/ptep/ptu081, german2024solutioncosmologicalobservablesstarobinsky} which are functions of these slow roll parameters.

%and restricting the parameter space using \textit{Planck} data -- $n_{s}=0.9623\pm0.0071$,  $10^{3}\alpha_{s} =-1.9\pm9.1$, and $r<0.034$. 

% \begin{figure}[h]
%     \centering
% \includegraphics[width=\linewidth]{New_analogy_diagram_1to1.pdf}
%     \caption{\blue{Keep the aspect ratio same as before.}}
%     \label{fig:Potential_analogy}
% \end{figure}
\begin{figure}[t]
    \centering
\includegraphics[width=\linewidth]{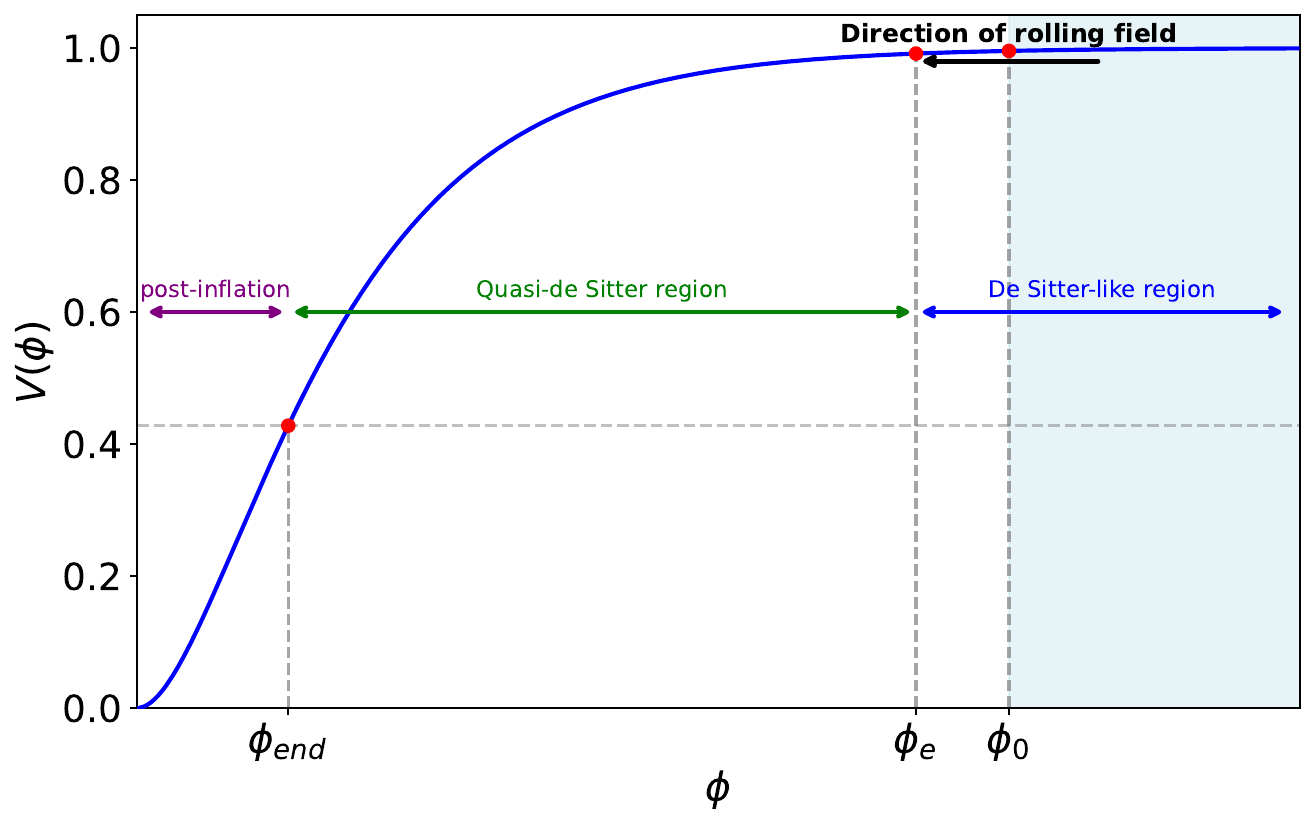}
\caption{Illustration of  inflationary potential $V(\phi)$ dynamics, the field $\phi$ evolves from right to left, beginning in an observationally inaccessible regime (light blue shaded region) before entering the observable CMB window $[\phi_0, \phi_e]$. The de-Sitterization condition in Eq. \eqref{eq:de-sitter-condition} has to be satisfied in both these regions. Finally, the field enters a quasi-de Sitter phase in the interval $[\phi_e, \phi_\mathrm{end}]$ prior to the end of inflation.}
% Dynamics of the inflationary potential $V(\phi)$ near the CMB window onset ($\phi_\mathrm{0}$). In this diagram, $\phi$ increases from right to left.
 % In this diagram, time increases from right to left. 
% The light blue shaded region represents observationally inaccessible  regime, considering the possibility that inflation may have set in before what observations allow us to probe. Interval between $\phi_0$ and $\phi_\mathrm{e}$ is the de Sitter epoch. de Sitterization condition \eqref{eq:de-sitter-condition} holds only during this period. Between $\phi_\mathrm{e}$ and $\phi_\mathrm{end}$, we have quasi de-Sitter period.}
%Towards left The intervals defined by small displacements $\phi_{e}$ delineate the  observationally accessible regimes, with the arrow indicating the direction of the rolling field. \red{. bridging the general, observationally inaccessible epoch and the observationally accessible slow-roll regime. . After that $\alpha$ may not satisfy this condition.}

\label{fig:Potential_analogy}
\end{figure}

\subsection{Constraints based on de Sitterization bounds\label{sec:bounds-on-parametrs}}
In \S \ref{sec:de-Sitter-with-potential}, we discussed that we can achieve the de Sitter state with the constant part of the inflaton potential without employing cosmological constant. But we also need to have $\phi$ dependent part in order to avoid  unphysical eternal inflation without a graceful exit.  %\red{[early-Universe dynamics are driven by inflation, which consists of two distinct phases: an initial de Sitter phase followed by a quasi-de Sitter phase~\cite{PhysRevD.107.023527}.]} But the initial period of inflation is de Sitter. In this section, we introduce a `de-Sitterization condition' and %establish its corresponding parameter bounds it entails.  
Following Ref. \cite{PhysRevD.107.023527, Bhatt2024InflationDynamics, mu2026inflationdrivenbarecosmological}, we divide the inflationary period into two phases (see Fig. \ref{fig:Potential_analogy})
\begin{enumerate}
\item \textit{de Sitter epoch} captures the early dynamics of inflation when the potential is  dominated by a constant term.  This \textit{initial} epoch in principle may start from an infinite time in the past {\cite{PhysRevD.28.2118,Kitada:1992uh,chakraborty2001inflation}}. Nonetheless, observable part of this epoch lies between $\phi_0$ and $\phi_\mathrm{e}$, where  $\phi_0$ is the scalar field corresponding to  the earliest time  accessible to CMB observations and $\phi_\mathrm{e}$ is the scalar field after $1e$ fold of $\phi_0$. {This % chosen $\phi_\mathrm{e}$ after $1e$ fold of $\phi_0$ 
is consistent with the fact that de Sitterization of an initial Bianchi geometry is induced within approximately one $e$-fold  \cite{kitada1992cosmic}.} It is the interval, $[\phi_0,\phi_\mathrm{e}]$, that allows us to put bounds on de Sitterization parameter $\alpha$  for existing inflationary potentials.  %$\phi\in[\phi_\mathrm{I},\phi_\mathrm{I}-\delta]$ which corresponds to an approximate de Sitter phase driven by a nearly constant potential.
\item \textit{Quasi de Sitter epoch} happens between $[\phi_\mathrm{e},\phi_\mathrm{end}]$ and driven by $f(\phi)$. % where it starts giving significant contribution. 
Identifying the plateau height $V_0$ as an effective initial cosmological constant ($  V_0\simeq\Lambda  $), the condition \eqref{eq:de-sitter-condition}  delineates the transition from pure de Sitter to quasi-de Sitter evolution. %\blue{For $V_0 = 0$, exact de Sitter expansion strictly requires Eq.~\eqref{eq:de-sitter-condition for v=0}; violating it transitions the system from approximate to quasi-de Sitter evolution.} \red{Do we really need this?}% and the evolution becomes quasi-de Sitter.
\end{enumerate}
%While standard slow-roll parameters are strictly constrained to the observable interval, we argue that sufficiently small field displacements leave the local dynamics approximately invariant. This justifies extending the slow-roll approximation albeit with highly suppressed values into the observationally inaccessible regime. Consequently, the potential parameters $A, B, C,$ and $D$ evolve according to these extended slow-roll dynamics. Substituting these into \eqref{eq:independent equation} yields the local coefficients $f_i$ ($i=0, \dots, 4$). Finally, we apply the constraints from \eqref{eq:bound model independend} to compute the minimum $\alpha$ that simultaneously satisfies all inequality bounds.
%anchor the analysis at $\phi=\phi_0$, the observationally accessible onset of the CMB window. Over the next single $e$-fold, the inflaton evolves to $\phi_\mathrm{e}$. %We analytically extrapolate backward into the unobservable prior epoch spanning $ \phi\in[\phi_\mathrm{I}+\delta,\phi_\mathrm{I}]  $. 
 % Within this domain we distinguish two dynamical phases  
We can relate initial and final scalar fields $\phi_\mathrm{f}$ and $\phi_\mathrm{i}$ values for a given potential based on the number of $e$-folds that have passed
\begin{equation}
\int_{\phi_\mathrm{f}}^{\phi_\mathrm{i}} d\phi \frac{V(\phi)}{V'(\phi)}=N \label{eq:efolds-scalar-field}
\end{equation}
%\red{As an illustrative example, we consider the observationally accessible de Sitter regime arising from the general solution given in Appendix~\ref{eq:defi-h-and-w}, restricting attention to the field interval $(\phi_0,\phi_e)$.}   
E.g., (i) $N=1$, if we consider $\phi_\mathrm{f}= \phi_\mathrm{e}$ and $\phi_\mathrm{i} = \phi_0$ and (ii) $N=60$, for $\phi_\mathrm{f}= \phi_\mathrm{end}$ and $\phi_\mathrm{i} = \phi_0$.
%Having established asymptotic de Sitterization under the condition \eqref{eq:de-sitter-condition}, 
Now we proceed towards establishing bounds on the parameter $\alpha$ based on given inflation model. For that, first  we write {$F(\phi)$ (c.f. \eqref{eq:de-sitter-condition})} corresponding to higher order terms of potential $V(\phi)$ as
\begin{equation}
    F(\phi) = \sum_{i= 1}^nc_i \phi^i \label{eq:F-expan}
\end{equation}
Our objective now is to put bounds on the coefficients $c_i$. Also, if inflation model is available, we can calculate bounds on $\alpha$. %To achieve this, we apply Theorem \ref{thm:extremal-Chebyshev-Property} which is applicable on polynomials defined over the interval $[-1,1]$. Thus to apply the result in the present context, we first need to map $\phi_0\to -1$ and  $ \phi_\mathrm{e}\to 1$ which is achieved  via \textit{affine transformation}. 
Details are given in Appendix \ref{sec:math-result} where we give the result for $n$.
As an illustration, setting $n=4$ in expression \eqref{eq:main-inequality} gives the following bounds. We choose $n=4$ to connect our theoretical results with the Planck observations, where slow-roll parameters up to fourth order are reported \cite{2020}. 
%gives the following $\phi$ independent bounds on  coefficients $f_i$ %in terms of parameters $h$ and $w$ are found to be
\begin{subequations}\label{eq:bound model independend}
\begin{align}
h^4|c_4|  &\le 8\alpha \\
|h^3|\ |c_3 + 4 c_4 w|  & \le 4\alpha \\
 h^2 |c_2 + 3 c_3 w + 6 c_4 w^2|  &\le 2\alpha \\
|h|\ |c_1 + 2 c_2 w + 3 c_3 w^2 + 4 c_4 w^3|  &\le \alpha \\
|c_1 w + c_2 w^2 + c_3 w^3 + c_4 w^4| &\le  \alpha
\end{align}
\end{subequations}
here $h= (\phi_\mathrm{e}-\phi_0)/2$ and $w=(\phi_\mathrm{e}+\phi_0)/2$. These relations can be `inverted' to put bounds on the expansion coefficients $c_i$. Using \eqref{eq:main-bound-cond} and $n=4$ then gives 
%gives
%Consequently, the extremal bounds on this localized feature are governed by shifted Chebyshev polynomials \cite{Porter:2005cu}. %After completing the above analysis, Applying minimax property \eqref{eq: main equation in inequality} of Chebyshev polynomials of the first kind on \eqref{eq:independent equation} gives the following $\phi$ independent bounds %we impose a condition that allows us to derive results from \ref{eq:independent equation}  independent of $\phi$, given by
% \begin{align}\label{eq:bound model independend}
% |f_4| \leq 8\alpha \hspace{0.5cm},|f_3| \leq 4\alpha, \hspace{0.5cm} |f_2| \leq 8\alpha, \hspace{0.5cm}|f_1| \leq 3\alpha
% \end{align}
%{We cannot make a definitive statement about this bound; however, we can assert that $f_{0} < \alpha$, or at least that $f(x) \leq \alpha$, given the way $f(x)$ is defined.}  Using these bounds, with the aid of triangle inequality gives
%Following this procedure, we obtain sharp upper bounds on the absolute values of the local coefficients. Translating these results to the scalar field framework, we map the local coefficients \ref{eq:bound model independend} onto the original coefficients of the potential using Triangle inequality, yielding
% which then gives  \blue{say something that $w>0$ we don't need to put $|\cdot|$}
\begin{subequations}\label{eq:original_parameter_bound}
\begin{align}
|c_4| &\leq \frac{8\alpha}{h^4} \\
|c_3| &\leq \alpha\left(\frac{4}{|h^3|}
      + \frac{32|w|}{|h^4|}\right) \\
|c_2| &\leq \alpha\left( \frac{2}{|h^2|}
      + \frac{12\,w}{|h^3|}
      + \frac{144\,|w|^2}{|h^4|}\right) \\
|c_1| &\leq \alpha\left(\frac{1}{h}
      + \frac{4|w|}{h^2}
      + \frac{36|w|^2}{h^3}
      + \frac{416|w|^3}{h^4}\right)
\end{align}
\end{subequations}
These constraints establish absolute limits on the de Sitterization mechanism for a general potential. We derive both model-independent \eqref{eq:bound model independend} and model-dependent \eqref{eq:original_parameter_bound} bounds, which remain strictly valid across the de Sitterization epoch %entire domain $\phi \in [\phi_1, \phi_2]$ or $x\in[-1,1]$  
provided the condition $|F(\phi)| \leq \alpha$ holds. %\blue{condition Eq.~\eqref{eq:de-sitter-condition}-\eqref{eq:de-sitter-condition for v=0} $|F(\phi)| \leq \alpha$ and $|f(\phi)| \leq \alpha$ holds.} %In this letter, we are calculating $\alpha$ within a precisely defined domain which we  call the \textit{initial} part of inflation. Because $V(\phi)$ is continuous, the bound $|f(\phi)|\leq\alpha$ (with $  \alpha<V_0  $) that enforces de Sitterization in the plateau region must hold throughout the entire interval. Our construction %strengthens \blue{is consistent with} this result because the same bound guarantees that an initially anisotropic Bianchi Universe reaches exact de Sitter expansion after precisely one $  e  $-fold under the potentials considered here. This framework directly addresses the inflationary initial-condition problem~\cite{Linde:2017pwt,Bhatt2024InflationDynamics} and imposes sharp, observationally anchored constraints on model parameters. Although current data probe only the initial de-sitter and  quasi-de Sitter regime, our analytic bounds translate into strict mathematical restrictions on the early, observable de Sitter epoch. We follow Linde \cite{Linde:2017pwt} by normalizing the potential to $V(\phi) \simeq 1$. Setting $V_0 = 1$ naturally restricts our $\alpha$  parameter to the strict bound $\alpha < 1$.
We demonstrate that this condition acts as a universal initial constraint at the onset of inflation. % any viable potential $V(\phi)$ must initially satisfy this de-Sitterization criterion to remain consistent with current Planck observational data. 
{Our analysis shows that a single field slow roll inflation model isn't viable if $\alpha>1$ during any point within this range.}

\subsection{A worked out example\label{sec:example}}
Recasting the potential $V(\phi)$ in terms of standard slow-roll parameters satisfying slow roll conditions, $|\epsilon_V|\ll1$, $|\eta_V| \ll 1$, etc., \cite{Sivaram:2007yd,2020} gives
\begin{equation}\label{eq: expanding potential compair}
\frac{V(\phi)}{V_{0}} = 1+\sqrt{2\epsilon_{V}}\,\phi+\frac{\eta_{V}}{2}\phi^{2}+\frac{\xi_{V}^{2}}{6\sqrt{2\epsilon_{V}}}\phi^{3}+\frac{\bar{\omega}^{3}_{V}}{48\epsilon_{V}}\,\phi^{4}
\end{equation}
%Imposing the slow-roll conditions $|\epsilon_V|\ll1$ and $|\eta_V| \ll 1$ (similar conditions hold for other parameters as well). that $\xi_V^2$ appears, and that $\bar{\omega}_V^3 \ll 1$ holds true as well. We recast the potential in terms of these parameters \cite{Sivaram:2007yd,2020}. The requisite flatness of $V(\phi)$ ensures that higher-order terms remain strictly subdominant to the constant scale {$V(\phi) \approx V_0$} throughout the observable epoch (initial regime). %Exploiting this quasi-de Sitter \blue{[shouldn't this be de Sitter?]} \red{No, because we use the term effective constant or approximately constant, we require this type of potential to realize a quasi-de Sitter phase.} behaviour where the potential acts as an effective constant, we derive stringent new upper bounds on the expansion parameters.
% \begin{equation}\label{eq: expanding potential compair}
% V(\phi)=V_{0}\left(1+\sqrt{2\epsilon_{V}}\,\phi+\frac{\eta_{V}}{2}\phi^{2}+\frac{\xi_{V}^{2}}{6\sqrt{2\epsilon_{V}}}\phi^{3}+\frac{\bar{\omega}^{3}_{V}}{48\epsilon_{V}}\,\phi^{4}\right)
% \end{equation}
Expanding till 4th order in $\phi$ in \eqref{eq:F-expan} and then comparing \eqref{eq:modified potential form} with \eqref{eq: expanding potential compair} gives 
\begin{equation}
c_1 = \sqrt{2\epsilon_{V}},\ c_2= \frac{\eta_{V}}{2},\ c_3= \frac{\xi_{V}^{2}}{6\sqrt{2\epsilon_{V}}},\ c_4 = \frac{\bar{\omega}^{3}_{V}}{48\epsilon_{V}}
\label{eq:relating-local-part-with-model}
\end{equation}
%The relationship which allows the parameters $A$, $B$, $C$, and $D$ to be determined. It is also worth noting that in another study \cite{Sarkar2025}, the authors proposed an alternative potential and examined its expansion, obtaining results that are consistent with observational data. Hence, this potential can likewise be regarded as a viable candidate for the test potential. Now, we evaluate the four slow-roll potential parameters, and the resulting values are then analysed in light of the observational constraints reported by the \textit{Planck} Collaboration \cite{2020}. To evaluate de Sitterization parameter $\alpha$, we select an arbitrarily small value from  interval specified for slow roll parameters, as per \eqref{eq:slow-roll-params-value}.
% \begin{subequations}
% \begin{align}
%     \epsilon_{V}&<0.0048 \\
%     \eta_{V}&=-0.0082 \\
%     \xi_{V}^{2}&=-0.004 \\
%     \bar{\omega}^{3}_{V}&=0.0048
% \end{align} 
% \end{subequations}
Next, we calculate $\alpha$, along the following steps. 
\subsubsection{Case 1: $V_{0}\neq0$}\label{sub:example1}
Next we take an example for $V_0\ne 0$ case and choose  poly-$\alpha$ attractor potential \cite{Kallosh_2022}
\begin{equation}
    V(\phi) = V_0\left(\frac{\phi^{6}}{\phi^{6}+(0.1)^{6}}\right)
\end{equation}
with $p=3$ and $\mu= 0.1$.
\begin{enumerate}
\item To simplify the calculations, we normalize the potential by setting  $V_0 = 1$.
\item The end point of inflation is characterized by $\epsilon_V=1$. From this, we can calculate scalar field at the end of the inflation, $\phi_\mathrm{end}$
\begin{equation}
    1 = \frac{1}{2}\left( \frac{V'}{V}\right)^2 \Rightarrow \phi_\mathrm{end} = 0.220
\end{equation}
\item Integrating \eqref{eq:efolds-scalar-field} for the given  potential gives 
\begin{equation}
    \phi_\mathrm{i}^2 = 4N + \phi_\mathrm{f}^2 \label{eq:integrated-efolds-eqn}
\end{equation}
The starting ($\phi_\mathrm{0}$) and end ($\phi_\mathrm{end}$) points  of inflation are $60e$ folds apart, see Fig \ref{fig:Potential_analogy}. For $N=60$, \eqref{eq:integrated-efolds-eqn} with $\phi_\mathrm{f} =\phi_\mathrm{end}$ gives 
$\phi_0\approx0.980 $, similarly, $N = 1$ corresponds to $\phi_\mathrm{e} \approx 0.976$.  
\item With $\phi_1= \phi_0$ and $\phi_2 = \phi_{\mathrm{e}}$, we obtain $h$ and $w$ of \eqref{eq:defi-h-and-w}. These are then used in \eqref{eq:bound model independend}, together with values of $c_i$'s obtained from \eqref{eq:relating-local-part-with-model} and \eqref{eq:slow-roll-params-value}.
\item All five conditions in \eqref{eq:bound model independend} are satisfied simultaneously if $\alpha \leq 1.02\times10^{-1}$ when $\epsilon_V = 0.0044$.
\end{enumerate}
In this case, our criterion gets satisfied. This model is also favoured by Planck data.
\subsubsection{Case 2: $V_0=0$}\label{sub:example2}
We choose quadratic potential {$V(\phi) = \phi^2$}.
\begin{enumerate}
\item We initially set $V_0 = 0$, there is no need to normalize the above potential.
\item The end point of inflation is characterized by $\epsilon_V=1$. From this, we can calculate scalar field at the end of the inflation, $\phi_\mathrm{end}$%,  can bei.e., . The end point is characterized by , and thus with the given potential
\begin{equation}
    1 = \frac{1}{2}\left( \frac{V'}{V}\right)^2 \Rightarrow \phi_\mathrm{end} = \sqrt{2}
\end{equation}
\item Integrating \eqref{eq:efolds-scalar-field} for the given  potential gives 
\begin{equation}
    \phi_\mathrm{i}^2 = 4N + \phi_\mathrm{f}^2 \label{eq:integrated-efolds-eqn1}
\end{equation}
The starting ($\phi_\mathrm{0}$) and end ($\phi_\mathrm{end}$) points  of inflation are $60e$ folds apart, see Fig \ref{fig:Potential_analogy}. For $N=60$, \eqref{eq:integrated-efolds-eqn1} with $\phi_\mathrm{f} =\phi_\mathrm{end}$ gives %an application of \eqref{eq:efolds-scalar-field} for the given potential gives 
$\phi_0\approx 15.56$, similarly $N = 1$ corresponds to  $\phi_\mathrm{e} \approx 15.43$. %Also scalar field after $1e$ fold which is also the point de Sitter epoch ends is given by $\phi_{e} = 15.43$.
\item With $\phi_1= \phi_0$ and $\phi_2 = \phi_{\mathrm{e}}$, we obtain $h$ and $w$ of \eqref{eq:defi-h-and-w}. These are then used in \eqref{eq:bound model independend}, together with values of $c_i$'s obtained from \eqref{eq:relating-local-part-with-model} and \eqref{eq:slow-roll-params-value}.
\item All five conditions in \eqref{eq:bound model independend} are satisfied simultaneously if $\alpha \ge 1.28\times10^{3}$ when $\epsilon_V = 0.0044$.
\end{enumerate}
The potential $V(\phi)$ thus contradicts the condition $\alpha< 1$ and hence our criterion says that this model is to be ruled out by observations.
%\blue{as shown in Examples \ref{sub:example1} and \ref{sub:example2}, the model exhibits the observed behaviour and satisfies the condition \(\alpha<1\).}
%Setting $V_0 = 1$ \blue{(this needs an explanation!)}, and using \eqref{eq:relating-local-part-with-model} in \eqref{eq:bound model independend} to express the bounds via standard slow-roll parameters. By evaluating the field evolution over the initial $e$-fold (from $\phi_{0}$ to $\phi_{e}$) and exploiting the smallness of the slow-roll parameters, we extract the minimal de Sitterization parameter $\alpha$ satisfying \eqref{eq:bound model independend}. Remarkably, evaluating this single parameter provides a robust constraint on candidate models. 
Some other cases are summarized in Table \ref{tab:tuning_parameter_models}. We can see that models successfully undergoing initial de Sitterization ($\alpha < 1$) are precisely those favoured by \textit{Planck} constraints. Consequently, the condition $\alpha < 1$ serves as a universal mathematical diagnostic for viable inflationary potentials, even though it cannot uniquely pinpoint the exact primordial potential driving the expansion. 

We must also point out that although in this example and all those given in Table \ref{tab:tuning_parameter_models}, we have considered slow roll parameters to determine $\alpha$. But if we can somehow determine the scalar field values $\phi_0$ and $\phi_\mathrm{e}$, then our formalism doesn't need any slow roll parameters and just based on the de-Sitterization condition \eqref{eq:alpha-cond}, we can constrain models of inflation.
\blue{\begin{table*}[htbp]
\centering
\small
\renewcommand{\arraystretch}{1.2}
\begin{tabular}{l l l l}
\toprule
\textbf{Viable Models} & $\alpha$ value & \textbf{Non-viable Models} & $\alpha$ value \\
\midrule
Hilltop Quadratic                          & $1.34 \times 10^{-1}$ & $\phi^{1/3}$                                  & $3.45 \times 10^{1}$ \\
D-Brane ($p=2$)                            & $2.16 \times 10^{-1}$ & $\phi^{2/3}$                                  & $1.39 \times 10^{2}$ \\
D-Brane ($p=4$)                            & $6.92 \times 10^{-2}$ & $\phi$                                        & $3.16 \times 10^{2}$ \\
T-$\alpha$-attractor ($m=2$, $\alpha=0.05$)                            & $8.56 \times 10^{-1}$ & $\phi^{2}$                                    & $1.28 \times 10^{3}$ \\
Double Well ($p=1$, $\phi_0=10$)                     & $4.97 \times 10^{-2}$ & $\phi^{3}$                                    & $2.92 \times 10^{3}$ \\
D-Brane ($p=6$)                            & $4.46 \times 10^{-2}$ & $\phi^{4}$                                    & $5.25 \times 10^{3}$ \\
D-Brane ($p=8$)                            & $3.42 \times 10^{-2}$ & Natural Inflation                             & $1.79 \times 10^{3}$ \\
T-$\alpha$-attractor ($m=3$, $\alpha=0.05$)                            & $9.99 \times 10^{-1}$ & Hilltop Quartic                               & $0.517\times 10^1$ \\
T-$\alpha$-attractor ($m=1$, $\alpha=0.05$)                  & $6.52 \times 10^{-1}$ & Double Well                                   & $0.229 \times 10^1$ \\
Mutated Hilltop ($\tau=3$)                 & $7.77 \times 10^{-1}$ & Exp. Tails                                    & $0.669\times 10^1$ \\
KKLT ($n=4$, $m=0.5$)                      & $5.65 \times 10^{-1}$ & Double Well ($p=3$, $\phi_0=0.01$)           & $5.15\times 10^4$ \\
Woods-Saxon ($b=0.001$, $c=-5$)            & $7.37 \times 10^{-3}$ & $\phi^{1/2}$                                  & $7.80 \times 10^{1}$ \\
Woods-Saxon ($b=0.001$, $c=-8$)            & $1.55 \times 10^{-2}$ & $\phi^{3/2}$                                  & $7.17 \times 10^{2}$ \\
Poly $\alpha$-attractor ($p=3$, $\mu=0.1$) & $1.02 \times 10^{-1}$ & $\phi^{5/3}$                                  & $8.87 \times 10^{2}$ \\
E-$\alpha$-attractor ($\alpha=0.1$)     & $7.70 \times 10^{-1}$ & Natural ($f=5$)                               & $0.187\times 10^1$ \\
E-$\alpha$-attractor ($\alpha=0.01$)    & $1.08 \times 10^{-1}$ & Natural ($f=7$)                               & $9.68 \times 10^{1}$ \\
Pseudo Natural                             & $2.34 \times 10^{-4}$ & Hilltop ($p=3$, $\mu=10$)                     & $0.193\times 10^1$ \\
KKLT ($n=2$, $m=0.05$)                     & $1.12\times 10^{-1}$ & $\phi^{4/3}$                                  & $5.65 \times 10^{2}$ \\
Loop Inflation ($\alpha=0.03$)             &  $4.07\times10^{-1}$  & Hilltop ($p=6$, $\mu=10$)                     & $1.38 \times 10^{1}$ \\
$^{*}$Starobinsky ($R+R^2$)                      & $1.92 \times 10^{1}$    & Exp. Tails ($q=2$)                            & $0.138\times 10^1$ \\
                                           &                       & T-model ($m=1$, $\alpha=10$)                       & $3.46 \times 10^{2}$ \\
                                           &                       & T-model ($m=2$, $\alpha=10$)                       & $7.78 \times 10^{2}$ \\
                                           &                       & Mutated Hilltop ($\tau=1$)                    & $1.22 \times 10^{1}$ \\
                                           &                       & KKLT ($n=2$, $m=5$)                           & $1.68 \times 10^{2}$ \\
                                           &                       & KKLT ($n=4$, $m=5$)                           & $1.80 \times 10^{2}$ \\
                                           &                       & Poly $\alpha$-attractor ($p=3$, $\mu=1$)      & $0.527\times 10^1$ \\
                                           &                       & Loop Inflation                                & $0.282\times 10^1$ \\
                                           &                       & Mixed Large Field                             & $4.71 \times 10^{3}$ \\
                                           &                       & Rad. Corr. Quartic                            & $5.22 \times 10^{3}$ \\
                                           &                       & Renorm. Inflection Point                      & $1.17 \times 10^{2}$ \\
                                           &                       & Pure Arctan                                   & $1.39 \times 10^{1}$ \\
                                           &                       & Fibre Inflation                               & $2.97 \times 10^{1}$ \\
                                           &                       & Hyperbolic ($n=1$)                            & $4.67 \times 10^{2}$ \\
                                           &                       & Hyperbolic ($n=2$)                            & $2.73 \times 10^{3}$ \\
                                           &                       &  Double Well ($p=2$, $\phi_0=10$)             & $0.159\times 10^1$ \\
\bottomrule
\end{tabular}
\caption{Table showing calculated $\alpha$ values for different models of inflation. Here `viable' models are the ones allowed by Planck data. It can be seen that all viable models except Starobinsky (shown with an asterisk) satisfy the de Sitterization criterion \eqref{eq:alpha-cond}. For non-polynomials models, $\alpha$ is calculated by terminating $V(\phi)$ Taylor expansion at 4th order.
}
\label{tab:tuning_parameter_models}
\end{table*}}

\section{Conclusion and Discussion}
We now give a summary of main results of this paper
\begin{itemize}
\item We have shown that an initially anisotropic Bianchi universe, driven by a single homogeneous and isotropic scalar field with a general potential $V(\phi)$, can dynamically evolve toward an approximate de Sitter phase. In contrast to scenarios that require a cosmological constant like potential, our result follows from imposing {the de Sitterisation condition, \eqref{eq:de-sitter-condition}}. The de Sitter phase emerges after a short intermediate epoch during which anisotropy decays.
\item We further showed, that for the potential \eqref{eq:modified potential form}, the higher-order contribution  remains subdominant when the overall bounds \eqref{eq:bound model independend} satisfy \eqref{eq:alpha-cond}. In this regime, the potential is effectively constant, allowing the de-Sitterisation condition to be realized for a broad class of scalar-field potentials. Importantly, the relevant criterion is not the bound (\ref{eq:original_parameter_bound}) on each individual {coefficient of $F(\phi)$}, but rather the bound on the function as a whole. Thus, \eqref{eq:alpha-cond} % condition $\alpha<1$ therefore 
provides a % alternative  simple and sufficient 
criterion for achieving de-Sitterisation with a general potential. 
\item As illustrated in Fig.~\ref{fig:Potential_analogy}, the relevant field range overlaps with the slow-roll interval \((\phi_0,\phi_e)\), where the condition \(\alpha<1\) holds. This overlap provides a direct connection between the early-time theoretical bound and the observationally allowed inflationary regime. In particular, Table~\ref{tab:tuning_parameter_models} shows that this simple bound excludes the same models disfavoured by \textit{Planck} data. Out of all the cases considered, we find only Starobinsky model as an exception. It doesn't satisfy our criterion, yet it is favoured by Planck data. Thus, we conclude that \eqref{eq:alpha-cond} serves as an alternative      
   %  compact and effective 
criterion for identifying  model having  actually gone through de-Sitterisation. 
%\red{\item Finally, the function coefficient bounds derived in Appendix~\ref{sec:math-result} are not limited to cosmological applications. They provide a general framework for constructing approximately constant functions coefficient bound and can be readily applied to a broad class of problems where such behaviour is required. } 
\end{itemize}
Our results suggest that de-Sitterisation is relevant only when the initial geometry is generalized away from the exactly isotropic and homogeneous case. For a fully isotropic and homogeneous , this mechanism is less significant. By contrast, in an inhomogeneous setting, the time scale and the bound may change and can depend on the parameters \(h\) and \(w\). This motivates further study of de-Sitterisation in more general cosmological backgrounds. We also believe that it is possible to constrain models of inflation just on the basis of determining $\alpha$ without the need of slow roll parameters. We plan to pursue this in future. Finally, we must state that the results of this paper are based on the following two assumptions
\begin{enumerate}
    \item The background evolution condition \eqref{eq:pure part} contains $\le$ sign. In order to solve for the perturbation \eqref{eq:perturbation_part}, we assumed only `$=$' sign holds. We are unable to explain how `$<$' sign can be ruled out.
    \item The de Sitter condition in \eqref{eq:de-sitter-condition} implies $V(\phi)\le V_0(1+\alpha)$. This means during de Sitter epoch $V(\phi)$ can take any value  but its upper bound has to satisfy $V_0(1+\alpha)$. In \eqref{eq:almost-constant potential}, we chose  $V(\phi)\approx V_0(1+\alpha)$ in order to establish lower bound on $K$.
\end{enumerate}
At this point of time, we are unable to justify these assumptions but we plan to study the assumptions and their ramifications in future.

\section{Acknowledgement}
Rahul Kothari acknowledges computing facilities availed through the IIT Mandi Grants IITM/SG/DIS-ROS-SPA/111 and IITM/SG/SS-DDP-KH-RKO/123. We also express our sincere gratitude to Subenoy Chakraborty, Bikash Chandra Paul, Sampat Sharma and  Syed Abbas for their valuable discussions and constructive suggestions. We are also very thankful to
the anonymous referee, whose comments helped us considerably improve the quality of the paper.

% \section{appendix}
\appendix
\section{Solving extrinsic curvature ODE beyond constant potential\label{sec: linear}}
%Ref. \cite{Heusler:1991ep} demonstrated that the de Sitter state cannot be realized for a general potential. 
In this appendix, we solve \eqref{eq:fin-diff-equa} perturbatively and show that we can realize an approximate de Sitter state due to $f(\phi)$ term in the potential $V(\phi)$. The departure from the pure de Sitter state is controlled by $\alpha$.  %Inspired by this finding, we therefore argue that, When initial geometric conditions are relaxed, a de Sitterization mechanism is essential to drive the system asymptotically toward a state that matches with late-time observations. This mechanism fundamentally relies on a nearly constant potential energy scale. Starting from \eqref{eq:fin-diff-equa}, We consider the case of a time-dependent potential that includes a non-linear contribution \eqref{eq:modified potential form}. 
We assume that \eqref{eq:fin-diff-equa} admits the following perturbative solution% the general solution to the differential equation  can be expressed perturbatively as
\begin{equation}\label{eq:perturbative_solution}
    K(t) = K_{0}(t) + p(t)
\end{equation}
Here, $K_{0}(t)$ denotes the background solution corresponding to the constant potential $V_{0}$ and thus a perpetual de-Sitter evolution, while $p(t)$ represents the first-order perturbative correction. %Higher-order corrections are neglected, since their contributions are assumed to be subdominant compared with the leading perturbative term.  
For a constant potential, the background evolution using \eqref{eq:fin-diff-equa} is found to satisfy
\begin{equation}\label{eq:pure part}
    \dot{K}_{0}(t) \leq -\frac{1}{3}K_{0}^{2}(t) + \kappa^{2}V_{0}.
\end{equation}
Solving this, following the method of \cite{c680dbbe-119c-31e0-a97a-d790b679674f} we recover the result reported in \cite{PhysRevD.28.2118}.
%Consequently, the upper bound on the solution of Eq.~\eqref{eq:fin-diff-equa} takes the form
\begin{equation}\label{eq:simplify-upper-bound}
    K_{0}(t)\leq\frac{3\zeta}{\tanh(\zeta t)}, \quad \zeta:= \kappa\sqrt{\frac{V_{0}}{3}}
\end{equation}
%After solving the constant-potential case, 
{Thus, we can interpret the constant part $V_0$ of the potential $V(\phi)$ as cosmological constant.}
We now generalize this result by substituting  \eqref{eq:perturbative_solution} and potential $V(\phi)$ (c.f. \eqref{eq:modified potential form}) in \eqref{eq:fin-diff-equa} which yields 
\begin{equation}\label{eq:perturbation_part}
     \dot{K}_{0}(t) + \dot{p}(t) \leq \kappa^{2}(V_{0} + f(\phi)) -\frac{K_{0}^{2}(t)}{3} - \frac{2}{3}K_{0}(t)p(t).
\end{equation}
%since $p(t)$ is assumed to remain sufficiently small, making its quadratic contribution subleading. 
Here, we've neglected the sub-leading $p^{2}$ term. Notice that \eqref{eq:perturbation_part} contains an inequality, so in order to solve this,  we proceed with the equality part of \eqref{eq:pure part}
\begin{equation}
    \dot{K}_{0}(t) = -\frac{1}{3}K_{0}^{2}(t) + \kappa^{2}V_{0} \Rightarrow K_{0}(t) = \frac{3\zeta}{\tanh(\zeta t)}\label{eq:assumption-2}
\end{equation}
to obtain the following equation satisfied by $p(t)$
\begin{equation}
\dot{p}(t)+\frac{2}{3}p(t) \frac{3\zeta}{\tanh(\zeta t)}\leq \kappa^2 f(\phi) \le\kappa^{2}V_{0}\alpha 
\end{equation}
where the last part is obtained using de-Sitterization condition. Solving this gives, %This has the form of a linear first-order differential inequality. Solving it, we obtain
\begin{equation}\label{eq:per solution}
p(t)\leq \frac{3\alpha }{2} \left[\frac{\zeta}{\tanh(\zeta t)}
-\frac{t\zeta^2}{\sinh^{2}(\zeta t)}\right].
\end{equation}
Now using \eqref{eq:assumption-2} and \eqref{eq:per solution} in  \eqref{eq:perturbative_solution} gives the following upper bound on $K(t)$
\begin{equation}\label{eq:total solution}
K(t)\leq \frac{3\zeta}{\tanh(\zeta t)}
+\frac{3\alpha}{2} \left[\frac{\zeta}{\tanh(\zeta t)}
-\frac{t\zeta^2}{\sinh^{2}(\zeta t)}\right]
\end{equation}
Corresponding lower bound on $K$ is obtained using  \eqref{eq:Guass-Codazzi-simpli} 
\begin{equation}\label{eq:inequality_from_7}
 \frac{K^{2}}{3} \geq \kappa^{2}V(\phi)
\end{equation}
Next we notice that this equation is to be solved during de-Sitter epoch during which potential is approximately a constant. Consider the case $V_{0}\neq0$ \eqref{eq:de-sitter-condition}, the potential always satisfies an upper bound $V(\phi) \le V_0(1+\alpha)$. We assume that potential approximately equals the upper bound  %implies $V(\phi)$  satisfies an . As $\phi$ varies, so does $V(\phi)$ but the bound always remains satisfied. But as the potential is approximately a constant during de-Sitter epoch, we write% As we stated before,  during the de Sitter epoch, the potential is roughly a constant. Thus we approximate% Based on the regime-specific potential behaviour established in Secs. \ref{sec:de-Sitter-with-potential} and \ref{sec:bounds-on-parametrs}, Based on those argument, we express the potential as
% Rather than applying  condition \eqref{eq:de-sitter-condition} across the entire cosmic expansion history, we focus  specifically on the de Sitter regime to derive the bounds on $K$. As we discussed before, pure de Sitter evolution requires a constant potential which is physically unrealistic. This situation is ameliorated by putting non-constant part which we can control using e \eqref{eq:de-sitter-condition}. Thus during the this epoch, we set 
\begin{equation}\label{eq:almost-constant potential}
    V(\phi)\approx V_{0}(1+\alpha) 
\end{equation}
Finally, using \eqref{eq:almost-constant potential} in \eqref{eq:inequality_from_7}, gives
%Thus, the potential remains approximately constant during the de Sitter phase. 
%We can obtain background solution% To make further progress, we substitute (i) the equality part of \eqref{eq:pure part} in   \eqref{eq:perturbation_part} and (ii) %for the equality case of Eq.  rather than the inequality ($<$) case and also % using the equality limits of Eq. \eqref{eq:pure part} along with the potential \eqref{eq:almost-constant potential}, which gives} We can now obtain the complete upper bound using \eqref{eq:per solution} in \eqref{eq:perturbative_solution} , and will serve as the basis for the subsequent analysis. % we obtain the following inequality,
% \begin{equation}\label{eq:more_simple}
%     \frac{K^{2}}{3}-\kappa^{2}V(\phi)\geq 0
% \end{equation}
% \red{Imposing the approximately constant potential \eqref{eq:almost-constant potential} on Eq. \eqref{eq:more_simple}, we obtain}
\begin{equation}\label{eq:below range}
    K\geq 3\zeta \sqrt{1+\alpha} \approx 3\zeta \left(1 + \frac{\alpha}{2}\right)
\end{equation}
where we have ignored the terms of $\mathcal{O}(\alpha^2)$. This is justified since as per \eqref{eq:per solution}, $p(t)\sim \alpha$ and we have neglected all terms beyond linear order in $p(t)$.
\section{Constraining $\sigma$ and $T_{ab}^M u^au^b$}
To find relevant bounds, we again utilize Gauss Codazzi relation \eqref{eq:Guass-Codazzi-simpli} which gives
\begin{equation}
    \frac{K^{2}}{3} - \kappa^{2}V(\phi) \ge
\frac{\sigma_{ab}\sigma^{ab}}{2} + \frac{-R^{(3)}}{2}
+\kappa^{2}T^{(M)}_{ab}u^{a}u^{b}.
\end{equation}
Now we use (i) the upper bound \eqref{eq:total solution} (ii) the approximation \eqref{eq:almost-constant potential} and (iii)  the result that if $A_i\ge 0$ for $i=1,2,3$ and $\sum_{i=1}^3A_i \le \epsilon$, then $A_i\le \epsilon$ for $i=1,2,3$. This gives 
% that if \(A,B,C\geq0\), the bound \(A+B+C\leq\epsilon\) enforces each  term to satisfy \(A,B,C\leq\epsilon\), we get
\begin{align}
0\leq\frac{\sigma_{ab}\sigma^{ab}}{2}&\leq \frac{3\zeta^{2}}{\sinh^{2}(\zeta t)}\left(1+ \alpha -\frac{t\alpha\zeta}{\tanh(\zeta t)}\right) \notag \\
&\quad +\frac{3\alpha^{2}\zeta^2}{4}\left(\frac{1}{\tanh(\zeta t)}-\frac{\zeta^{2}}{\sinh^{2}(\zeta t)}\right)^2 \label{eq:upperbound of sigma}
\end{align}
% \begin{widetext}
% \begin{align}\label{eq:upperbound of sigma}
%     0\leq\frac{\sigma_{ab}\sigma^{ab}}{2}\leq \frac{3\zeta^{2}}{\sinh^{2}(\zeta t)}\left(1+ \alpha -\frac{t\alpha\zeta}{\tanh(\zeta t)}\right)+\frac{3\alpha^{2}\zeta^2}{4}\left(\frac{1}{\tanh(\zeta t)}-\frac{\zeta^{2}}{\sinh^{2}(\zeta t)}\right)^2
% \end{align}
% \end{widetext}
Similar upper bounds on the scalars (a) $T_{ab}^M u^au^b$ and (b) $-R^{(3)}/2$ are found to be
\begin{equation}
    0 \le \kappa^2 T^{(M)}_{ab}u^{a}u^{b}\leq  \mathrm{RHS},\  0 \le -\frac{R^{(3)}}{2}\leq  \mathrm{RHS}\label{upper bound of dominant part}
\end{equation}
where RHS is the expression following second inequality of \eqref{eq:upperbound of sigma}.

\section{Bounds on the coefficients of $F(\phi)$ \label{sec:math-result}}
In this section, we derive bounds on the coefficients of function $F(\phi)$ (c.f. \eqref{eq:F-expan}) that represents higher order contribution to the inflation potential $V(\phi)$. This conclusion rests on an important theorem,  which we state here without proof. For more details (including the proof), the reader is referred to Ref. \cite{tang2023quantum, cheney1982approximation}. %In this section, we state a theorem (without proof) that gives bounds on a polynomial $P_n(x)$ defined on the interval $[-1,1]$. , %and has been derived using tools such as the Chebyshev Alternation Theorem, de la Vallée Poussin theorem, and the extremal property of Chebyshev polynomials.
\begin{theorem}\label{thm:extremal-Chebyshev-Property}
Let
\begin{equation*}
    P_n(x)=\sum_{k=0}^{n} a_k x^k
\end{equation*}
be a polynomial of degree \(\leq n\) such that $|P_n(x)| \leq \alpha$ for some $\alpha>0$ and for all $x\in[-1,1]$.
% Write the Chebyshev polynomial \(T_k(x)\) in the monomial basis as
% \begin{equation*}
%     T_k(x)=\sum_{j=0}^{k} c_j^{(k)} x^j .
% \end{equation*}
Then we have an upper bound on the coefficients
\begin{equation*}
|a_k| \le 
\begin{cases}
     2^{k-1}\alpha & k\geq 1 \\
    \alpha & k=0
\end{cases}
\end{equation*}
for \(k=0,1,\dots,n\). 
% \begin{equation}\label{eq: main equation in inequality}
% |a_k| \le
% \begin{cases}
%      2^{k-1} & k\geq 1 \\
%     1 & k=0
% \end{cases}
% \end{equation}
\end{theorem}
As was discussed in \S\ref{sec:bounds-on-parametrs} that de Sitterization happens between a specific interval (see also Figure \ref{fig:Potential_analogy}). To keep the analysis general, we assume these to be   %The de Sitterisation phase occupies the intermediate interval 
$\phi_1$ and $\phi_2$. 
% $\phi_0$ and $\phi_\mathrm{e}$. 
In order to be able to apply Theorem \eqref{thm:extremal-Chebyshev-Property}, we map $\phi_1 \to -1$ and $\phi_\mathrm{2}\to 1$ which can be achieved via an affine transformation.
\begin{equation}
    \phi = xh+w,\text{ with }  h = \frac{\phi_\mathrm{2}- \phi_1}{2},\ w = \frac{\phi_1 + \phi_\mathrm{2}}{2} \label{eq:defi-h-and-w}
\end{equation}
In terms of $x$, we can write \eqref{eq:F-expan} as
\begin{equation}
    F(x) = \sum_{i=0}^n F_ix^i, \text{ with } F_i = h^i \sum_{k=i}^n c_k \binom{k}{i} w^{k-i}
\end{equation}
where $\binom{k}{i}$ denotes the binomial coefficient. Now an application of Theorem \eqref{thm:extremal-Chebyshev-Property} for $i\ge 1$ gives 
\begin{equation}
     \left| \sum_{k=i}^n c_k \binom{k}{i} w^{k-i} \right| \le \frac{2^{i-1}\alpha}{|h^i|} \label{eq:main-inequality}
\end{equation}
for $i=0$, we get the following inequality
\begin{equation}
    \left|\sum_{k=1}^n c_k w^{k}\right| \le \alpha
\end{equation}
From \eqref{eq:main-inequality}, we can get bounds on the coefficients $c_{n-i}$ using the triangle inequality
\begin{equation}
    |a-b|\ge ||a|-|b|| \ge |a| - |b|
\end{equation}
We can start with obtaining bounds on $c_n$ and can then work backwards to $c_1$ in the end. After some algebraic manipulations,  it can be shown that for $0\le i \le n-1$
\begin{align}
    |c_{n-i}| &\le \alpha \frac{2^{n-i-1}}{|h^{n-i}|} \Bigg[1 + \sum_{k=1}^i 2^k\left|\frac{w}{h}\right|^k \notag \\ &\quad \prod_{j=1}^{k} (n-i+j)
    \left(1 + \sum_{l=1}^{k-1}\frac{k-l}{(l+1)!}\right)  \Bigg] \label{eq:main-bound-cond}
\end{align}
where it is understood that if the upper index of the sum is smaller than the lower one, sum equals $0$.

\bibliographystyle{apsrev4-1}
\bibliography{reference}

\end{document}